\documentclass[prl, aps, 10pt, showkeys, twocolumn, nofootinbib]{revtex4-1}

\usepackage[utf8]{inputenc}
\usepackage[english]{babel}
\usepackage{amsmath}
\usepackage{amssymb}
\usepackage{graphicx}

\DeclareMathOperator{\Li}{Li}

\begin{document}

\title{The statistical description of the electron system on the liquid helium surface}

\author{B.\,I.~Lev}
\author{V.\,P.~Ostroukh}
\author{V.\,B.~Tymchyshyn}
\author{A.\,G.~Zagorodny}
\affiliation{Bogolyubov Institute for Theoretical Physics National Academy of Science,
Metrolohichna St. 14-b, Kyiv 03680, Ukraine}

\begin{abstract}
It is known that homogeneous distribution of particles in Coulomb-like systems can be unstable, and spatially inhomogeneous structures can be formed.
A simple method for describing such inhomogeneous systems and obtaining spacial distributions of electron density is proposed and applied to the case of two-dimensional electron systems on surface of liquid helium.
A free energy functional for the model in mean field approximation is obtained.
Creation of various types of structures, such as long-range periodical modulation and multi-electron dimples, is predicted by minimizing this functional.
\end{abstract}

\maketitle

\section*{Introduction}

Particle systems with Coulomb interaction (Coulomb-like systems), such as plasmas, colloidal particles, electrolyte solutions, electron on the helium surface etc., are widely presented both in nature and under laboratory conditions.
An interest to this system is generated by its applications to the studies of a variety of peculiar phenomena in various fields of science~\cite{fortov,lowen,edelman}.
Theoretical description of Coulomb like systems is one of important problems of statistical physics.
One of the problems here is a statistical description of Coulomb-like systems with high concentrations of interacting particles~\cite{lev_zagorodny}. 
Such effects as formation of different crystal structures, transitions between different phases are observed when concentration increases. 

Recently much interest has been generated to the experimental and theoretical studies of the low-dimensional Coulomb like systems.
Such systems are widely presented in experiments with emulsions, foams, polymers, colloidal suspensions etc.
Considerable attention is attracted to the special case of electrons on the surface of a dielectric substrate~\cite{cole,cole_cohen}.
Let us say, the first experimental realization of the Wigner solid, predicted in well-known article~\cite{wigner}, was made in an electron system on liquid helium~\cite{grimes_adams}.
Studies of these systems are not only of academic interest but can also have some practical applications.
For example, it is proposed to use the electrons on dielectric surface for quantum computations~\cite{platzman_dykman}.

A possibility of creating a two-dimensional system on the surface of a dielectric medium was predicted by~\cite{cole, cole_cohen, shikin}. 
First experiments were carried out one year later~\cite{williams_crandall_willis}. 
Two-dimensional electron systems are still extensively studied, and many interesting results have been obtained.
Let's now discuss some properties of these systems.
Electrons located on a dielectric surface have two degrees of freedom only~\cite{edelman,ando_fowler_stern}.
They can exist in forms of fluid or Wigner crystal~\cite{wigner,platzman_fukuyama,grimes_adams}.
Some interesting effects are caused by deformation instability of liquid helium surface, that causes, for example, phase transitions between triangle and square Wigner lattices~\cite{haque}.
Electrons on the surface of liquid helium become localized in macroscopic dimples when the electric field perpendicular to the surface exceeds a critical value.
These dimples form a two-dimensional hexagonal lattice \cite{leiderer1}.
Was investigated the phase transitions from a homogeneous two-dimensional charge distribution to the modulated charge density regime and observed the hysteresis effects that the transition is discontinuous \cite{ebner}.

Modern research in field of the low-dimensional electron systems is based mostly on the quantum field theory~\cite{zinn-justin} and the scaling theory~\cite{thouless}.
For example, electron transport properties in heterostructures and electron structures on the liquid-helium surface can be studied using quantum field theory methods~\cite{datta}.
As regards the scaling theory, it was worked out in~\cite{monarkha_kono}.
Nevertheless, these models are complicated for the analysis and require a lot of calculations.
Therefore, it would be highly desirable to introduce simpler quasi-classical models of the type~\cite{lev_zhugayevich,lev_zagorodny}, which could be efficient for the description of the properties of the low-dimensional electron systems.

Another interesting side of studying electrons on liquid helium is that this system is a representative of class of systems with long-range resembling Coulomb interaction. 
Dusty plasmas, systems of colloidal particles, electrolyte solutions significantly differ from each other by physical properties, but their inter-particle interaction causes formation of structures of one type, concerning the formation of stable periodical structures~ \cite{lev_zagorodny,leiderer2, shikin_leiderer,tsui_stormer_gossard,laughlin, GrainDynamics, StronglyCollisional,myPublication}.
Thus, we can expect that methods applicable to two-dimensional electron systems also can be useful for description of other systems of mentioned type.

The studies of low-dimensional electron systems have long history, but we do not have consistent statistical theory of such systems until now.
In particular, the thermodynamic conditions of structure formation are not yet known.
The main goal of the present contribution is to work out a model for the description of electrons on a helium surface in the presence of the external electric field applied perpendicularly to the surface and use it for the prediction of structure formation.
In order to do this we employ the concept of the effective interaction energy.
For the case of electrons on a cryogenic liquid substrate, distortion of the surface introduces additional physical effects~\cite{haque,shikin}. 

In the present contribution we study the influence of new  details of the electron interaction related to such effects.
We take into account both direct Coulomb repulsion and polarization interaction as well as the interaction due to the deformation of the helium surface~\cite{haque,shikin}.
Proposed simplified approach is suggested by the fact that the forces, governing self-organization, act on a length scale which is larger than the molecular size; as a consequence many specific details of the molecules of interest are not necessary for studying general features of phases.

We look at the electrons on the liquid helium surface in a way typical for the systems with long range interaction developing necessary formalism of  the statistical description of such systems in terms of the mean field approximation~\cite{lev_zagorodny,krasnoholovets_lev}.
The proposed statistical model makes it possible to describe the electron structure formation, especially long range periodical structure and many-electron dimple formation for various values of the electron density, temperature and the external electric field.
In this article we will discuss the surface instability itself, especially forming long range structures (comparing to Wigner crystal period) --- many-electron dimples~\cite{leiderer2,shikin_leiderer} and periodical structures~\cite{gorkov_chernikova}, that can be treated as Wigner crystal formed by these dimples.

The formation of clusters under thermodynamic equilibrium conditions was investigated theoretically \cite{lev_zhugayevich} as well as using computer simulation \cite{imperio_reatto, imperio_reatto2}.
In two-dimensional case this problem has been analyzed within the scenario of competing interactions: typically, a short-range attractive interaction against a repulsive short-range one \cite{archer, imperio_reatto, imperio_reatto2}.
This leads to unimportance of boundaries for structure formation in system.
But in case of Coulomb-like interaction they can be important due to their long-range nature.
So, boundaries are fully taken into account in our work and these effects are not lost.

Presented article is organized as follows. We begin from the discussion of general formalism for the statistical description of the equilibrium systems in the mean-field approximation.
The treatment we perform is rather general since we do not use the explicit form of the inter-particle potential and dimensionality of the system (Sec.~2).
In Sec.~3 we introduce continuous approximation, specify electron-electron potential and consider possible forms of the spatial electron distributions according to border conditions.
The results of the computer calculations considered in Sec.~4 show, that this term can cause significant change of the electron distribution function.
In conclusions we discuss the effects predicted in our treatment and the possibility to observe them in experiments.

\section{Statistical description of interacting particle systems}
\label{sec:stat_descr}

We begin with a brief calculation for inhomogeneous system of interacting particles~\cite{lev_zhugayevich}.
We wouldn't specify our system first with a view to obtaining universal theory that can be extended to other types of systems.
As we know, stable states of the system minimize its free energy, so our goal now is to find an expression for it.

In our model macroscopic states of the system will be described by a set of occupation numbers.
For a wide number of systems Hamiltonian can be written in form:
$$ 
  H(n) = \sum_s \epsilon_s n_s + \frac12 \sum_{s,s^{\prime}} V_{ss^{\prime}} n_s n_{s^{\prime}}.
$$
Here $\epsilon_s$ is the additive part of particle energy (usually it is kinetic energy, but also it can be an energy in external field.), $s$ indicates the particle states, $V_{ss^{\prime}}$ is interaction energy between particles in states $s$ and $s^{\prime}$, $n_s$ is the occupation number of state $s$.
Most of systems mentioned in introduction, including electrons on liquid helium surface, are essentially classical so we neglect any quantum correlations. 

Partition function of the system will be:
$$
  \begin{aligned}
    Z 
      &= \sum_{\{n_s\}} \exp (-\beta H) 
	= \sum_{\{n_s\}} \exp \Biggl[ 
	  - \beta \Biggl(
	      \sum_s \epsilon_s n_s + 
	  \Biggr.
	\Biggr. \\
      &+ \Biggl.
	 \Biggl.
              \frac{1}{2} \sum_{s,s^{\prime}} V_{s s^\prime} n_s n_{s^\prime}
        \Biggr) 
    \Biggr].
\end{aligned}
$$

In order to perform formal summation in this equation, we use the well-known properties of Gaussian integrals over auxiliary fields:
$$
\begin{aligned}
\exp \left( 
         \frac{\nu^2}{2 \vartheta} \sum_{s,s'} \omega_{ss'} n_s n_{s'} 
     \right) &= 
\int\limits_{-\infty}^{\infty} D\varphi \exp \Biggl( 
    \nu \sum_s n_s \varphi_s - \Biggr. \\
   &- \Biggl.
        \frac{\vartheta}{2} \sum_{s,s'} \omega_{ss'}^{-1} \varphi_s \varphi_{s'} 
    \Biggr),
\end{aligned}
$$
with 
$D \varphi = \frac{\prod_s d\varphi_s}{\sqrt{\det(2 \pi \beta \omega_{ss'})}}.$ 
We can get rid of quadratic dependence on occupation numbers of states, carrying it to introduced field.
Now we can write the partition function of such system as\footnote{In future infinite limits of integration will be omitted.}:
$$
Z = \int D\varphi \exp \left[
        \sum_s (i \varphi_s - \beta \epsilon_s) n_s -
           \frac{1}{2\beta} \sum_{s,s'} \left( 
           V_{ss'}^{-1} \varphi_s \varphi_{s'} 
       \right) 
    \right].
$$
Let us consider canonical ensemble. To fix the number of particles in the system we will use a Cauchy equation:
$$
\frac{1}{2 \pi i} \oint \xi^{\sum_s n_s - N - 1} d\xi = 1.
$$
We get partition function for N-particle system:
$$
\begin{aligned}
Z_N &= \frac{1}{2 \pi i} 
       \oint d\xi \int D\varphi \exp\Biggl[ 
          - \frac1{2\beta} \sum_{s,s'}
                V_{ss'}^{-1} \varphi_s \varphi_{s'} - 
       \Biggr.\\
    &  \Biggl. 
          - (N+1) \ln\xi 
       \Biggr] \prod_s \sum_{\{n_s\}} \left[
           \xi \exp(i \varphi_s - \beta \epsilon_s)
       \right]^{n_s}.
\end{aligned}
$$
We can perform summation by occupation numbers in obtained equation, according to type of statistics.
We get:
$$
Z_N = \frac1{2 \pi i} \oint d\xi \int D\varphi \exp [
        - \beta F(\varphi,\xi)],
$$
with an effective free energy:
\begin{equation}
\label{betaF0}
\begin{aligned}
  \beta F&(\varphi,\xi) 
= \frac1{2} \sum_{s,s'} V_{ss'}^{-1} \varphi_s \varphi_{s'} + \\
&+ \delta \sum_s \ln \left(
       1 - \delta \xi e^{-\beta \epsilon_s+i \varphi_s}
   \right) + (N+1) \ln \xi . 
\end{aligned}
\end{equation}
Variable $\delta$ indicates type of statistics under consideration.
It equals $+1$ for Bose-Einstein statistics, $0$ for Maxwell-Boltzmann statistics and $-1$ for Fermi-Dirac statistics.

We have obtained the expression for the free energy of the system of interacting particles in representation of auxiliary fields and chemical activity of particles $\xi = \exp ( \beta \mu )$.
It contains the same information as the original partition function with summation over the occupation numbers, i.e. all information about probable states of the system.
The partition function represented in terms of the functional integral over auxiliary fields corresponds to the construction of an equilibrium sequence of probable states of the system with regards to their weights.
Extension to the complex plane makes it possible to apply saddle-point method to find an asymptotic value of partition function, so we don't need to apply perturbation theory.
Dominant contribution is given by the states which satisfy the extremum condition for the functional:
$$
  \frac{\delta \beta F}{\delta \varphi}
= \frac{\delta \beta F}{\delta \xi} 
= 0.
$$
Variating~(\ref{betaF0}) we obtain the system for determination of saddle-point states:
\begin{subequations}
\begin{eqnarray}
  &\frac1{\beta} \sum_{s'} V_{ss'}^{-1} \varphi_{s'} 
- \cfrac{i \xi e^{-\beta\epsilon_s + i \varphi_s}}
       {1 - \delta\xi e^{-\beta\epsilon_s + i \varphi_s}} 
= 0; \label{saddle1}\\
  &\sum_s \cfrac{\xi e^{-\beta \epsilon_s + i \varphi_s}}
              {1 - \delta \xi e^{-\beta \epsilon_s + i \varphi_s}} 
= N+1.  \label{saddle3}
\end{eqnarray}
\end{subequations}
We can see from (\ref{saddle3}), that expression
\begin{equation}
  \label{f_s}
  f_s = \frac{\xi e^{-\beta\epsilon_s + i \varphi_s }}
           {1 - \delta\xi e^{-\beta\epsilon_s + i \varphi_s}}
\end{equation}
can be treated as averaged state occupation number.

Obtained system gives us possibility to get directly saddle-point states, that can be interpreted as thermodynamically stable distributions.
As we can see, there is an inverse interaction matrix in equation~(\ref{saddle1}).
Its determination for fixed interaction potential is a difficult mathematical problem itself.
It is known~\cite{edwards_lennard, samuel} that for potentials of the form $\omega_{ss^{\prime}} = \omega(|\vec{r}_s - \vec{r}^{\prime}_s|)$ in continuous 
case the inverse interaction matrix is given by:
$$
  \omega_{ss^{\prime}}^{-1} = \delta_{ss^{\prime}} \hat{L}_{r^{\prime}},
$$
where $\hat{L}_{r'}$ is the operator, for which the interaction potential is a Green function.
We know the inverse operator for the screened Coulomb potential~\cite{edwards_lennard, samuel, hubbard}:
\begin{equation}
  \hat{L}_{r'} = -\frac{1}{4 \pi q^2} (\Delta_{r'} - \lambda^2),
\end{equation}
where $q^2$ is interaction constant and $\lambda^{-1}$ is the screening length.
Problem for some particular potentials has been solved in~\cite{lev_zhugayevich}.

For more general interaction potential the inverse operator is unknown, so we have to find a way to turn back to direct operators.
Such possibility is given to us by~(\ref{f_s}).
Using this expression, we can perform inverse transformation to the occupation numbers:
$$
\begin{aligned}
&\varphi_s = i \beta \sum_{s'} V_{ss'} f_{s'},\\
&\frac1{2\beta} \sum_{s,s'} V_{ss'}^{-1} \varphi_s \varphi_{s'} 
  = - \frac{\beta}2 \sum_{s,s'} V_{ss'} f_s f_{s'},
\end{aligned}
$$
and rewrite the free energy:
\begin{equation}
\label{betaF1}
\begin{aligned}
 \beta F[f,\xi] 
= - \frac{\beta}{2} \sum_{s,s'} V_{ss'} f_s f_{s'} 
 &- \delta \sum_{s} \ln (1 + \delta f_s) \\
 &+ (N+1) \ln \xi (f).
\end{aligned}
\end{equation}
For canonical ensemble we should get rid of chemical potential in this equation. 
We can get it from~(\ref{f_s}):
$$
\ln\xi(f) \equiv \beta\mu = \beta(\epsilon_s + E_s) + \ln f_s - \ln (1 + \delta f_s).
$$
with
\begin{equation}
\label{mean_field}
E_s = \sum_s V_{ss'} f_{s'}.
\end{equation}
We consider chemical potential to be a constant for all the system, but it is useful for calculation to take it averaged over all the states:
$$
\ln \xi (f)
     = \frac{1}{N}\sum_s f_s \left[ 
           \beta(\epsilon_s + E_s) + \ln f_s - \ln (1 + \delta f_s) 
       \right].
$$
Now we can put this expression into the free energy~(\ref{betaF1}) and get:
\begin{equation}
\label{free_energy}
\begin{aligned}
   \beta F[f] 
&= \beta \sum_s f_s \epsilon_s 
 + \frac{\beta}{2} \sum_{s,s'} V_{ss'} f_s f_{s'} + \\
&+ \sum_s \left[f_s \ln f_s - (f_s+\delta) \ln (1 + \delta f_s) \right].
\end{aligned}
\end{equation}
This expression can be easily interpreted. 
First term is a kinetic energy of the system, second one is a potential energy.
The third term is a contribution of entropy, and it equals zero for $T=0.$

Now we can check the compliance of our theory to the classical results.
Let us now consider grand canonical ensemble.
For fixed chemical potential we can get from~(\ref{f_s})
generalization of the well-known distributions:
\begin{equation}
  f_s = \frac{1}{e^{\beta (\epsilon_s - \mu_s) } - \delta}.
  \label{rozp_uzag}
\end{equation}
with a generalized chemical potential:
$$
\mu_s = \mu - E_s.
$$
From this expression it is obvious that the saddle point approximation is equivalent to mean field approximation.
For ideal gas $\mu_s \equiv \mu,$ and we get classical statistical distributions.
To obtain distribution of particles in grand canonical ensemble, we can solve integral equation~(\ref{rozp_uzag}).
But for numerical calculation~(\ref{free_energy}) becomes more useful, and it will be applied in next sections of our article.
 
\section{Free energy continuous approximation}
\label{sec:cont_appr}

\subsection{Physical model of two-dimensional electron system on the liquid helium surface}
\label{subsec:model}

In this section we will apply proposed statistical description to the system of electrons on liquid helium surface.
It is well-known that electron is attracted to the surface of any dielectric media~\cite{landau_lifshitz_8}. 
But in addition to attractive part for electron-helium binding potential there is a repulsive part.
It comes from the fact that bulk liquid has no free place for additional electron because of being built from inert atoms with complete electron shells~\cite{skachko}.
This means that electrons are floating above the liquid helium surface.
It is sufficient that due to small dielectric constant for liquid helium distance between surface and electron layer is rather large (it is estimated as $76$\,\AA~\cite{edelman, koutsoumpos, lambert, skachko}), so we can neglect influence of separate atoms of surface.
So, we can treat the whole electron subsystem as two-dimensional~\cite{edelman, koutsoumpos, lambert, skachko}.
This point of view was experimentally illustrated in~\cite{brown_grimes}.
It was shown there that resonant frequency depends only on perpendicular component of applied magnetic field, and their mobility corresponds to two-dimensional electron liquid.
We will stay on this point of view and will neglect the third-dimension effects.
This statement should be reconsidered when taking into account helium surface deformations, but even in such case this approximation can be used by introducing effective attraction~\cite{haque}.

It was theoretically and experimentally shown, that electrons on the liquid helium surface can undergo a phase transition, which appears in their ordering stage~\cite{kosterlitz_thouless,grimes_adams} (liquid-to-solid transition).
This is not the only phase transition that can be observed.
In~\cite{haque} the structural transition of a Wigner lattice from triangle to square was investigated.
In~\cite{gorkov_chernikova} it was shown that homogeneous distribution of electron density is not always stable, and there are critical parameters when spacial structures, especially periodic deformations and multi-electron dimples, are formed.
Experimental confirmation for this can be found in~\cite{koutsoumpos}.
This structure formation can be regarded separately from Wigner crystallization, because characteristic length of such structures is much larger than the period of the Wigner lattice. 
In strong external fields electrons are collected and pushed to liquid helium film that causes creation of bubble-like structures called in different sources bubblons or dimples~\cite{edelman}.
They can include $\sim 10^4$ particles and can be treated as one quasi-particle with high effective mass.
Our goal is to find properties of these structures for different experimentally controllable parameters, especially temperature, electron density and electric field.

\subsection{Model potential}
\label{subsec:potential}

The main goal of this subsection is to introduce the explicit form of potential in order to use it in computer and analytic calculations.
According to~\cite{edelman, koutsoumpos, skachko} the electron potential in the presence of liquid helium film on the substrate is:
\begin{equation}
\label{eePotentialClassical}
\begin{aligned}
   V_{ee}(r) 
&= \frac{e^2}{4\pi\varepsilon_0 r} - \frac{4\Lambda_s}{\sqrt{r^2+(2d)^2}},\\
   \Lambda_s 
&= \frac{\varepsilon_s - \varepsilon_{He}}
        {16\pi\varepsilon_0(\varepsilon_s + \varepsilon_{He})}
   e^2.
\end{aligned}
\end{equation}
Here $\varepsilon_s$ and $\varepsilon_{He}$ are substrate and the liquid helium dielectric constants respectively, $r$\, is the distance between electrons, $d$\, is the helium film thickness.
The first term is related to the ordinary Coulomb interaction, the second one is the result of the liquid helium and substrate polarization.
These two terms describe direct interaction between electrons.
Notice, that the second term is different from that one used in~\cite{haque} and takes into account polarization of substrate and provides additional attraction, which creates ability for structure formation.
Really, an effect of surface-buckling instability can be interpreted using the idea of competition between the attractive and the repulsive interactions.
But polarization attraction is rather weak comparing to direct Coulomb interaction.
To enhance it we should involve another attractive force between electrons.
In our model it will be effective attraction produced by helium surface deformation.

In the presence of an external field electrons can be pressed against the helium surface with the force that exceeds gravitational one by many orders.
On the other hand, they cannot just go through this surface as being pushed out due to quantum effects~\cite{koutsoumpos, lambert, skachko, edelman}.
So it should be considered with imminence that the electrons can act on the liquid helium surface with the significant force.

In the introduction it was assumed that the system under consideration is two-dimensional.
But taking helium surface deformations into account "electron layer"\ deflection should be considered as well, which means that third dimension is also active and should in some way be included in calculation.
This problem is solved by adding to~(\ref{eePotentialClassical}) effective capillary interaction.
The lateral capillary interaction between two electrons on the helium surface was calculated in~\cite{haque}.
It can be presented in the form:
\begin{equation}
V_{\textrm{def}}(r) = - \frac{F^2}{2\pi\sigma}  \textrm{K}_0(\lambda r).
\label{deformPotential}
\end{equation}
Here $F=eE$ is the actual force that acts on each electron by the external field (in our case it is electrical clamping field), $\sigma$ is the surface tension of liquid helium, $r$ is the distance between particles, $\text{K}_0$ is the modified Bessel function, and $\lambda = 1/l_0,$ where $l_0$ is the capillary length that depends on the fluid properties only.
If external field is absent, surface deformations are caused only by weight of electrons and can be neglected, and field produces significant deformation and changes the energy of interaction between electrons. 
So, our model potential of electron-electron interaction will be:
\begin{equation}
\label{eePotential}
\begin{aligned}
   V(r) &= \frac{e^2}{4\pi\varepsilon_0 r} - \frac{4\Lambda_s}{\sqrt{r^2+(2d)^2}} - \frac{F^2}{2\pi\sigma}  \textrm{K}_0(\lambda r).\\
\end{aligned}
\end{equation}

\subsection{Free energy}
\label{subsec:free_energy}

In Sec.~2 we have developed general formalism for the free energy of quantum systems in mean field approximation.
Now, we will apply it to system of electrons on liquid helium surface.

Let us use the equation~(\ref{free_energy}) to obtain the free energy of such system.
For Fermi particles we use $\delta = -1.$ Dimensionality of the system $d=2.$  
We can use continual approach and write:
\begin{equation}
  \begin{aligned}
    &\beta F[\mu] = \int \frac{d^2 p\, d^2 r}{(2 \pi \hbar)^2} \frac{\beta p^2}{2 m} \frac{1}{e^{\beta(p^2 / (2 m) - \mu(\vec{r}))}+1} + \\
    & {} + \frac{\beta}{2} \iint \frac{d^2 p\, d^2 r}{(2 \pi \hbar)^2} \frac{d^2 p'\, d^2 r'}{(2 \pi \hbar)^2} V(|\vec{r}-\vec{r}'|) \times \\
    & {} \times \frac{1}{e^{\beta(p^2 / (2 m) - \mu(\vec{r}))}+1} \frac{1}{e^{\beta(p'^2 / (2 m) - \mu(\vec{r}'))}+1} + \\ 
    & {} + \int \frac{d^2 p\, d^2 r}{(2 \pi \hbar)^2} \left( \frac{1}{e^{\beta(p^2 / (2 m) - \mu(\vec{r}))}+1} \times \right. \\
    & {} \times \ln \frac{1}{e^{\beta(p^2 / (2 m) - \mu(\vec{r}))}+1} + \frac{1}{e^{-\beta(p^2 / (2 m) - \mu(\vec{r}))}+1} \times \\
    & {} \left. \times \ln \frac{1}{e^{-\beta(p^2 / (2 m) - \mu(\vec{r}))}+1} \right).
  \end{aligned}
\end{equation}

We can integrate this expression over the momentum.
It is necessary to notice that performing of such integration in the case of the Bose statistics will cause loss of the Bose-condensation effects.
In our case we do not have such problems and thus we get:
\begin{equation}
  \begin{aligned}
    &\beta F[\mu] = \frac{m^2}{8 \pi^2 \hbar^4 \beta} \int \int d^2 r\, d^2 r' V(|\vec{r}-\vec{r}'|) \times \\   
    & {} \times \ln(1+e^{\beta \mu(\vec{r})}) \ln(1+e^{\beta \mu(\vec{r}')}) + \frac{m}{2 \pi \hbar^2 \beta} \int d^2 r  \times \\   
    & {} \times \left[ \Li_2 \left( \frac{1}{e^{\beta \mu (\vec{r})}+1} \right) - \Li_2 \left( \frac{1}{e^{- \beta \mu (\vec{r})}+1} \right) - \right. \\
    & {} \left. - \Li_2 \left( -e^{-\beta \mu (\vec{r})} \right) - \frac{\pi^2}{6} \right].
  \end{aligned}
  \label{betaF_mu}
\end{equation}
Here $\Li_2$ is polylogarithm of the second order which is also called dilogarithm.
It is useful to introduce dimensional constant -- thermal length:
$$
  \lambda_T = \sqrt{2 \pi^2 \hbar^2 \beta / m}.
$$
Dispersion relation for electrons has the quadratic form:
\begin{equation}
  \varepsilon_s = \varepsilon(p) = p^2 / (2m).
\end{equation}
Of course it will be deformed in the presence of external potential, but this expression can be taken as the first approximation.
We treat the distribution of charge to be continuous in this system.
So electron density could be described by the function $\rho(x,y)$.
This assumption looks quite reasonable, since the electron gas under consideration is highly degenerated~\cite{edelman}. 

Finally, we should write the functional of the free energy in terms of electron density.
We can take from~(\ref{rozp_uzag}):
$$
  \rho(\vec{r}) = \int \frac{d^2 p}{(2 \pi \hbar)^2} \frac{1}{e^{\beta (p^2/(2m) - \mu(\vec{r}))}+1} = \frac{\pi}{\lambda_T^2} \ln (1 + e^{\beta \mu(\vec{r})}).
$$
Applying this to~(\ref{betaF_mu}), we have:
\begin{equation}
  \begin{aligned}
    &\beta F[\rho] = \frac{\beta}{2} \iint d^2 r\, d^2 r' V(|\vec{r}-\vec{r}'|) \rho(\vec{r}) \rho(\vec{r}') + \frac{\pi}{\lambda_T^2} \times \\
    & {} \times \int d^2 r \left[ \vphantom{\frac{e^{- \pi^{-1} \lambda_T^2 \rho(\vec{r})}}{1 - e^{- \pi^{-1} \lambda_T^2 \rho(\vec{r})}}} \Li_2 \left(e^{- \pi^{-1} \lambda_T^2 \rho(\vec{r})} \right) - \Li_2 \left(1 - e^{- \pi^{-1} \lambda_T^2 \rho(\vec{r})} \right) - \right. \\
    & \left. - \Li_2 \left(- \frac{e^{- \pi^{-1} \lambda_T^2 \rho(\vec{r})}}{1 - e^{- \pi^{-1} \lambda_T^2 \rho(\vec{r})}} \right) - \frac{\pi^2}{6} \right].
    \label{betaF}
  \end{aligned}
\end{equation}
As is seen, the free energy of our system consists of two parts: the potential energy of the system and the kinetic energy and contribution of the entropy. 
This functional can be easily used to determine thermodynamically stable states, that have to minimize this functional.
We obtain such states directly in the form of the electron density and no additional mathematical transformations are needed.

\subsection{Boundary conditions}
\label{subsec:boundary}

Till this point we still have not considered all the assumptions we have done.
Due to long-range nature of Coulomb-like forces boundary conditions may notably change the result, because particles start to ``feel'' size of the system.
Moreover, they introduce restrictions to the properties of the density function, that simplifies not only analytical, but computer calculations as well.
In~\cite{archer, imperio_reatto, imperio_reatto2} exponentially decreasing potential was chosen so boundaries could be ignored.
But for the system under consideration they should be taken into account.

We suppose electrons to be situated in a grounded square metal box with dimensions $-L...L$.
In most experiments ``box'' is actually round~\cite{skachko,koutsoumpos}, but it would make minimization more tricky and wouldn't introduce new physical effects.
Of course we can expect electrons to leave the system through the grounded walls.
But electron leakage through the boundaries is prevented by the guard ring or the guard potential~\cite{skachko}.
We assume that such guard field acts only on the electrons in the vicinity of the walls and can be neglected anywhere else~\cite{koutsoumpos}.

\begin{figure}[tbh]
\centering
\includegraphics[width=8cm,height=1.5cm]{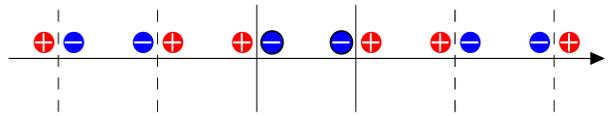}
\caption{
Symmetrical charges (blue with black edge) inside the box (walls are shown with a solid black line) over the liquid helium surface (shown with an arrow).
Due to the walls charge images are induced (shown without edge).
Besides, images of the walls are shown (dashed lines).
}
\label{fig:-1}
\end{figure}

Due to classical electrodynamics it can be shown that we should take into account imaginary charges along with the real ones (principle of images) when calculating total potential energy.
Nevertheless this causes inappropriateness of our calculations near the boundary, density disturbance we are interested in also appears far from the boundary as well.
To make our calculations simpler we will analytically continuate density distribution function to $(-\infty; \infty)$.
Moreover we claim it to be symmetrical (due to the system's symmetry) and antiperiodical (due to metal walls).
Figure~\ref{fig:-1} illustrates this two principles in a schematic way.
Every function that satisfies these conditions can be represented in a form of series:
\begin{equation}
\label{dfunction}
\rho(x;y) = \sum\limits^{\infty}_{i,j=0}
          C_{i,j}\,
          \cos\left(x\alpha_i\right)\,
          \cos\left(y\alpha_j\right),
\end{equation}
where $\alpha_i=({\pi}/{2L})[2i+1]$.
We leave to the reader prove of the fact that charge density $e\rho(x;y)$ in the form~(\ref{dfunction}) automatically holds principle of the symmetry ($e\rho(x;y)=e\rho(-x;y)=e\rho(x;-y)=e\rho(y;-x)$) and moreover when $|x|>L$ or $|y|>L$ $e\rho(x;y)$ can be treated as the value of the imaginary charge, for example $e\rho(L-x;y)=-e\rho(L+x;y)$.

Now we can also concretize potential energy part of free energy functional~(\ref{betaF}):
\begin{equation}
  F_{p} \approx \iint\limits_{-\infty}^{+\infty} d\chi d\gamma V(\chi;\gamma) \iint\limits_{-L}^{+L} dxdy \, \rho(x;y)\rho(x-\chi;y-\gamma).
  \label{eq:F_p}
\end{equation}

Here we have changed variables $x'\rightarrow\chi=x-x'$, $y'\rightarrow\gamma=y-y'$ and have taken into account that $V$ depends only on $|r - r'|$~(\ref{eePotential}).
Inner integral has boundaries from $-L$ to $+L$ because we are integrating all over the system.
On the other hand outer integral boundaries are set from $-\infty$ to $+\infty$ because of image charges.
Every electron has infinite number of images and calculating its energy makes us take them all into account.
Equality in this equation is approximate because of summand in $V$ that is not caused by electrostatical forces~(\ref{eePotential}).
We may notice that this summand is very ``short-range'', so we can consider integration with infinite boundaries as well.

Obtained equations~(\ref{dfunction}) and~(\ref{eq:F_p}) can significantly simplify following calculations.
If we substitute~(\ref{dfunction}) into potential energy expression~(\ref{eq:F_p}), expand by well known trigonometric equations $\cos\bigl([x-\chi]\alpha_i\bigr)$ and $\cos\bigl([y-\gamma]\alpha_j\bigr)$ and take into account
$$
\begin{aligned}
\int\limits_{-L}^{+L}\cos(x\alpha_i)\sin(x\alpha_j)dx&=0,\\
\int\limits_{-L}^{+L}\cos(x\alpha_i)\cos(x\alpha_j)dx&=\delta_{i;j},\\
\end{aligned}
$$
it can be shown
\begin{equation}
  \label{simpleChargeAction}
  F_{p} = \sum\limits^{\infty}_{i,j=0}C_{i;j}^2 k_{i;j},
\end{equation}
where $k_{i;j}$ depends on $\varepsilon_{He}$, $\varepsilon_{s}$, $L$ and $d$ only, and
$$
k_{i,j}=
\iint\limits_{-\infty}^{+\infty}
\cos(\chi\alpha_i)\cos(\gamma\alpha_j)V(\chi;\gamma)d\chi d\gamma.
$$
Also it can be shown that $k_{i;j}\geq0$.

When making computer calculations we can use the recurrence relations and precalculation of $k_{i,j}$ that significantly reduces the computation time.
It should be mentioned that in the case of canonical ensemble the charge of the system is constant.
So it should be:
\begin{equation}
\label{chargeLaw}
Q_{total} = \iint\limits_{-L}^{+L} \rho(\vec{r}) d^2 r.
\end{equation}
Taking explicit form  of $\rho(\vec{r})$ eq.~(\ref{dfunction}) into account it can be shown, that~(\ref{chargeLaw}) is of the form:
\begin{equation}
\label{simpleChargeLaw}
Q_{total} = \sum\limits^{\infty}_{i,j=0}C_{i;j} q_{i;j},
\end{equation}
where $q_{i;j}$ depend on $L$ only.

All subsequent calculations are made taking antiperiodical boundary conditions into account.
So obtained equations are very valuable as the core of following calculations.

\section{Minimization of free energy functional}
\label{sec:minimization}

\subsection{Analytic approach}

First let's make some general assumptions about effects that will be present in our system basing on free energy functional.
Let us consider simplified situation assuming that the temperature is equal to zero.
In~(\ref{free_energy}) we have three terms.
The third term automatically becomes equal to  zero at $T=0,$  since all electrons are  in their ground states and mean occupation numbers of states $f_s$ are 1 or 0.
The first term also can be neglected, because electron gas can be treated as degenerated~\cite{edelman}, so most of electrons have zero kinetic energy.
Thus, for low temperatures potential energy itself defines the configuration of the ground state of electron layer, and we can minimize only it.
Similar situation is observed when densities of electrons are high, because the potential energy $F_p \propto \langle \rho \rangle^2,$ and other terms depend on the density only linearly.

We should notice that if we neglect all other terms as mentioned above and minimize $F_p$, we can try to find result analytically.
Suppose, that instead of infinity we are summing up to some big enough $N$.
If we will change variables $C_{i;j}\rightarrow c_{i;j}=C_{i;j}/\sqrt{k_{i;j}}$, then it turns out to be a problem of finding point, where $N$-dimensional sphere eq.~(\ref{simpleChargeAction}) touches $N$-dimensional plane eq.~(\ref{simpleChargeLaw}).
But in this case charge distribution has nonconstant sign that is nonphysically.
This means that sphere radius should be bigger and result we want to obtain is one of the points of of hypersphere and hyperplane intersection.
To find this point or points we should use numerical methods, or somehow introduce additional restrictions.

\subsection{Triangular vs square lattice comparison}

In the previous subsection we tried to get the electron distribution function analyzing our model potential in some simplified situation with antiperiodical boundary conditions.
Result we have got was non-physical because our minimization procedure didn't involve sufficient restrictions for distribution function e.g. it's non-negativity inside of ``metal box''.
In this section we will try to get along with analytics by choosing some additional restrictions for the electron density distribution function.
Of course method developed in this section can be doubted as the one involving too much intuition-based assumptions but it can be applied to other systems with rather different model potential and make some at least qualitative predictions.
So it seems to be quite rational to give at least an outline for this approach.

It is known from~\cite{wigner} that electrons tend to form a triangle lattice, and the same behavior can be expected for the system of charged and interacting dimples.
On the other hand, in~\cite{haque} it was shown, that a phase transition between triangle and square lattice exists in system of electrons under consideration.
So, it is interesting to know preferred lattice type for system of electron dimples on liquid helium surface.

To answer this question we should calculate free energy of triangular and square lattice and compare them.
First task is to obtain analytic expression for both lattices' electron density.
We don't know real charge distribution in one dimple, so we should approximate it with some function.
We have chosen Gaussian distribution of charge because of its simplicity and conformity to real dimple form:
$$
G(x) = \frac{1}{\sigma\sqrt{2\pi}} e^{-\cfrac{x^2}{2\sigma^2}}.
$$
With this distribution we can obtain free energy of these lattices in analytic way.

\begin{figure}[tbh]
\centering
\includegraphics[width=8cm,height=5cm]{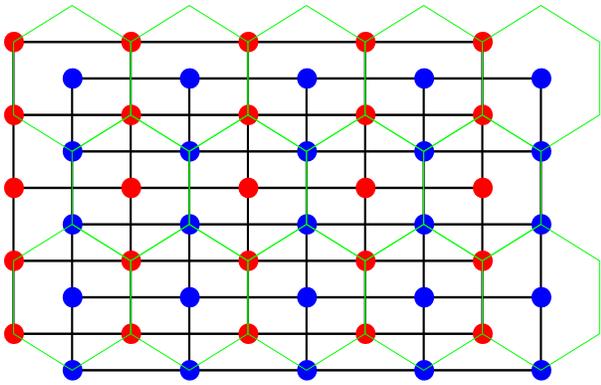}
\caption{Triangular lattice decomposed into two rectangular lattices.}
\label{fig:Lattice}
\end{figure}

As the first step we should find the Fourier decomposition of this two lattices.
Let us assume that we have $M$ Gaussian peaks on the $(-L \dots L)$ interval.
Then with some simplifications and assuming that $\sigma$ is small compared to $L/M$ the last equation can be rewritten as:
$$
G_M(x) = \frac{M}{2L} 
            \sum_{n=-\infty}^{\infty} 
                e^{-\cfrac{M^2n^2\pi^2\sigma^2}{2L^2}}
                \cos\left( \frac{x M n \pi}{L} \right)
$$
To write all equations in more compact way we can use the notation:
$$
\gamma = \frac{\pi^2\sigma^2}{8L^2}
$$
Obtained Gaussian decomposition is still not suitable for calculations because it violates principle of the image charges.
To make it correct we can decompose a unit step function and multiply it by obtained decomposition of the Gaussian function.
After rearrangement, simplification and assuming that $\sigma$ is small compared to $L/M$ we can get:
$$
\begin{aligned}
\mathcal{G}_M(x) &=
             \sum_{m = 0}^{\infty} 
             \frac{(-1)^m}{L}\sin^{-1}\left(\frac{\pi(1+2m)}{2M}\right) \times \\ &\times
             e^{-\gamma(2m+1)^2}
             \cos\left(x \alpha_{m}\right)
\end{aligned}
$$
At this point we want reader to pay attention to similarity of last equation and~(\ref{dfunction}).
Square lattice can be represented as $\mathcal{G}_M(x)\mathcal{G}_M(y)$.
Triangular lattice can be decomposed into two rectangular lattices~(Fig.~\ref{fig:Lattice}) and represented as $\mathcal{G}_K(x)\mathcal{G}_N(y)+\mathcal{G}_K(x+L/K)\mathcal{G}_N(y+L/N)$.
Here $K = [N\sqrt{3}]$ (by square braces we mean integer part of the argument) and $2KN=M^2$ (mean charge densities are equal for both lattices).
This representations can be treated as partial cases of~(\ref{dfunction}).

Using equations~(\ref{eePotential}) and~(\ref{betaF}) we can calculate energy for this lattices and compared it.
Physical parameters were taken as follows \cite{koutsoumpos}: $L=1\,cm$; $\varepsilon_s=11$ (Silicium); $d=0.1\,cm$.
We will find global minimum of free energy calculating it for different number of peaks and lattice types.

\begin{figure}[tbh]
\centering
\includegraphics[width=10cm,height=8.9cm]{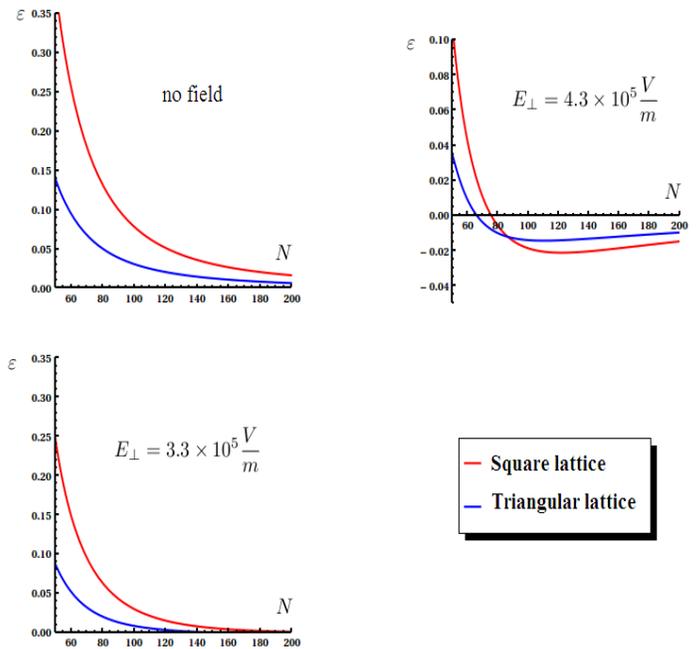}
\caption{Free energy of different lattices. \\ 
$T=0\,K$, energy $E=\varepsilon Q^2/L$, $Q$\,is the total charge, $N$\,is the number of peaks along one dimension
in square lattice, total number of peaks is $N^2$. Dispersion of one ``gaussian`` is $\sigma=10^{-3}$.}
\label{fig:0}
\end{figure}

As we can see from Fig.~\ref{fig:0}, really phase transition between triangular and square lattice exists.
For low pressing fields triangular lattice is favorable from the point of view of free energy.
But when we make pressing field strong enough this situation changes and square lattice begins to realize its global minimum.
We suppose that this phase transition can be observed in experiments.

\subsection{Numeric minimization}

Previously we tried to treat all calculations analytically, but due to high complexity of the equations it eather needs a lot of assumptions or results in non-physical distribution.
This subsection we will use computer calculations to provide most sufficient way of describing system under consideration.
Moreover it allows us to treat temperature as nonzero in this subsection.

Now some explicit form of $F$ can be written. 
$C_{i;j}$ can be treated as variables and then $F$ can be minimized with gradient descent method. 
This method was chosen as most suitable due to its simplicity and low calculation complexity that makes possible to work with big amount of variables $C_{i;j}$.
Since we will not get into technical details for those readers unfamiliar with gradient descent method we recommend to refer to appropriate literature (for example~\cite{nesterov}).
But it should be mentioned that there are some features when using gradient descent method in this paper. 

Method can find values of the variables when some function reaches it's minimum.
But here we should find when not function rather a functional is minimal.
Decomposing $\rho(\vec{r})$ into Fourier series and substituting into the functional we reduce the problem to finding coefficients of the Fourier series.
Of course it is infinite number of these coefficients, so we should take only first $N$.
The more we take the more precise the result will be after minimization procedure (in this paper $10^4$ coefficients were taken).
Though, we have physical limitation: in one period of the lowest term of series we should have significant number of electrons to be able to use continuous approximation.
Physical parameters were taken as follows~\cite{koutsoumpos}: $L=1\,\text{cm}$; $\varepsilon_s=11$ (Silicium); $d=0.1\,\text{cm}$. 
The second feature is that $\rho$ should be nonnegative so sign of the $\rho(x;y)$ is checked on every step of the gradient descent method.
Due to this calculations we will get some values of the coefficients $C_{i;j}$ that minimize free energy $F$.

\begin{figure}[tbh]
\centering
\includegraphics[width=8cm,height=4cm]{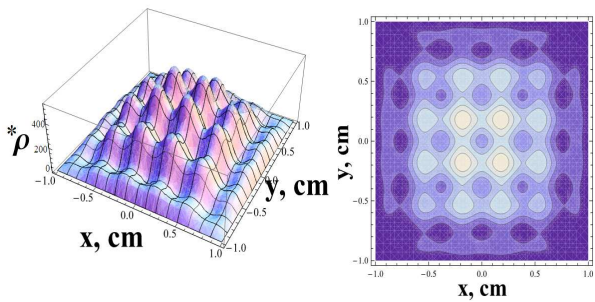}
\caption{The plot of the electron distribution function. \\ $T=0.02\,K$, $\strut^*\!\rho=\rho L^2/Q_{total}\cdot100\%$, no external field.}
\label{fig:1}
\end{figure}

Without an external field, electrons form  2D periodical structure if the temperature is low enough~(Fig.~\ref{fig:1}).
This corresponds to the case of the Wigner crystal, formed by dimples.
Lighter areas correspond to higher electron concentration (higher probability of finding some electron there).
This means that stable state of two-dimensional electron system on liquid helium surface is not simple electron Wigner crystal, but also some periodical structures, that can be interpreted as periodical modulations of local density or another Wigner crystal formed by dimples.
At this point we want reader to pay attention to presented result scale.
Presented figures clearly show some long-wave inhomogeneities in electron distribution, but short-wave and small disturbances can be hidden so it doesn't violate previous section results.
If we suppress long-wave disturbances in the obtained distribution function we will notice that previous section may be treated as some approximate description of it.

\begin{figure}[tbh]
\centering
\includegraphics[width=8cm,height=4cm]{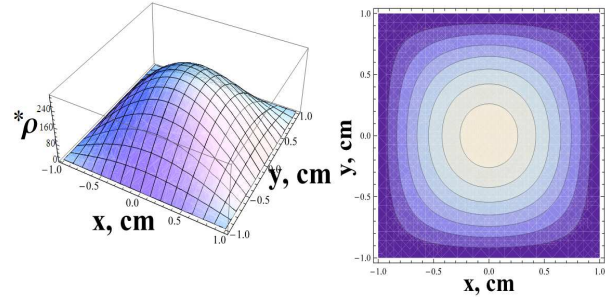}
\caption{The plot of the electron distribution function.\\ $T=0.4\,K$, $\strut^*\!\rho=\rho L^2/Q_{total}\cdot100\%$, no external field.}
\label{fig:2}
\end{figure}

Temperature growth makes possible to observe some smearing of electron density function that confirms our intuition~(Fig.~\ref{fig:2}).
Due to grounded boundaries, particle distribution function is always zero there, so we cannot expect uniform distribution function to be a constant, but it tends to be as temperature and computation precision grow.
But such effect is noticeable only in the case of low electron densities.
When densities are high, electrostatic interaction fully prevails, and structures are formed are equivalent to low temperature ones up to scaling of plot.
 
\begin{figure}[tbh]
\centering
\includegraphics[width=8cm,height=4cm]{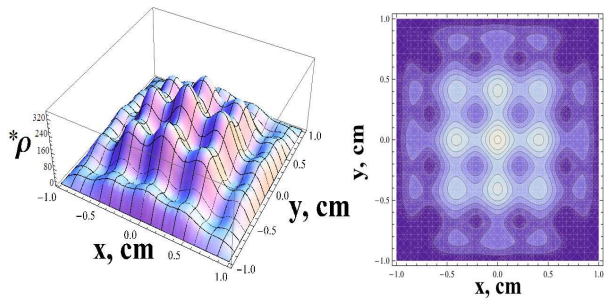}
\caption{The plot of the electron distribution function. \\ $T=0.02\,K$, $E=1.4 \cdot 10^5\, V/m$, $\strut^*\!\rho=\rho L^2/Q_{total}\cdot100\%$.}
\label{fig:3}
\end{figure}

Turning an external field on can only deepen the difference between local maximum and minimum of electron distribution function because of increase of rivalry between attraction and repulsion.
According to our model, electrons should collect together due to effective attraction caused by the helium film deformations.
Computer simulation results again confirm the intuition and show that some sharpening of density function can be observed~(Fig.~\ref{fig:3}). 
Of course zones with higher particle concentration will be more deformed as zones with lower particles concentration.

\begin{figure}[tbh]
\centering
\includegraphics[width=8cm,height=4cm]{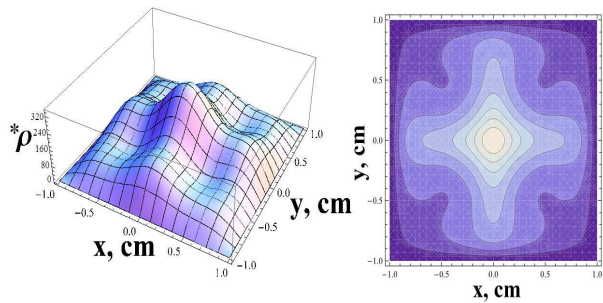}
\caption{ The plot of the electron distribution function.\\  $T=0.02\,K$, $E=2 \cdot 10^5\, V/m$, $\strut^*\!\rho=\rho L^2/Q_{total}\cdot100\%$.}
\label{fig:4}
\end{figure}

Further enhancement of holding electric field will cause some more significant changes of surface structure.
Due to growth of attraction characteristic size of structures will also grow up.
(Fig.~\ref{fig:3})~shows that the local maximum and minimum are additionally shifted.
Also there is a tendency of gathering to more and more large structures.
If electron concentration is low enough, electrons can gather into one large dimple in helium film~(Fig.~\ref{fig:4}).
In that case attraction prevails over repulsion, so distribution again becomes less fragmented, but more sharp than in the case of~(Fig.~\ref{fig:2}), where density tends to be constant.
On the other hand, when the density is big enough, they can be pushed inside helium and form some kind of bubble full of electrons (bubblon).
Of course such behavior cannot be taken into account with so simple model, but some appropriate estimates can be done.

\section{Conclusions}
\label{sec:conclusions}

We purposed two aims while working on this paper.
First aim was to develop general formalism for obtaining particle distributions in systems of charged particles, that can describe collective behavior of system, but rather simple to work with it comparing to quantum field and renormalization theories.
For this we have obtained a free energy expression for system of particles in mean field approximation, that depends only on two-particle interaction potential.
It is suitable for arbitrary systems, that satisfy two conditions.
First is that we can neglect two-particle correlations, that is a criteria of ability to use mean field approximation, that is equivalent to saddle-point approximation.
Second comes from the ability of Bose-condensation: to take it into account, we can't simply integrate by momentum with distribution, as we have done in~(\ref{betaF_mu}), but generally this model can be adapted to this case.
If our system satisfies this, problem of obtaining its particle distribution reduces to minimizing of free energy functional with proper boundary conditions.
Our model is simple  to be solved without using clusters or other powerful computer tools and with some additional simplifications can be even studied analytically. 

Second aim was to analyze behavior of two-dimensional electron systems on liquid helium surface in electric fields and explain some experimental facts, that can be found for example in~\cite{koutsoumpos}.
We achieved this using model developed in first part of our article~(Sec.~2).
We used effective interaction potential between electrons, that takes surface curvature into account.
This potential can be used for low densities of electrons, when multi-particle terms of interaction decomposition can be neglected, and for low gradients of surface to be able to perform series expansion by it~\cite{haque}.
This means that this expression is very good for low densities and low pressing electric fields, but its accuracy decreases if some of this is wrong.
More precise definition of it will need a great mathematical work.

It was shown that ground state of the electron layer is inhomogeneous.
We came to a conclusion that presence of the "metal walls" and their form can significantly influence electron distribution function inducing the periodical structures in it.
With the temperature grows we can see smearing of the distribution function and it tends to be the uniform one when the temperature is high.
We have also show the importance of the external electric field for the multi-electron dimples formation and found that the surface deformation leads to essential changes in the properties of the structures.
In particular, holding electric field can cause collection of electrons to bigger structures, and in strong fields electrons tend to gather into one great dimple. 
If the field is weak, then only local maximum of electron distribution function can be observed.
This explains the results, observed experimentally~\cite{koutsoumpos}, in particular separation of the electron layer into two components which interact with magnetic field in different manner.
Modern experimental technique allows to measure electron distribution functions on the liquid helium surface, so our results could be compared with experimental data.

We are grateful for the joint grant of Russian Foundation for basic research and National Academy of Science of Ukraine.


\begin{thebibliography}{99}
\bibitem{fortov} V.\,E.\,Fortov, A.\,V.\,Ivlev, S.\,A.\,Khrapak, et al., Phys. Rep. 421, 1 (2005).
\bibitem{lowen} H.\,Lowen, Phys. Rep.,{\bf 237}, 249 (1994).
\bibitem{edelman} V.\,S.\,Edelman, Uspekhi Fizicheskikh Nauk {\bf 23}, 227 (1980).
\bibitem{lev_zagorodny} B.\,I.\,Lev and A.\,G.\,Zagorodny, Phys. Rev. E {\bf 84}, 061115 (2011).
\bibitem{cole} M.\,W.\,Cole, Phys. Rev. B {\bf 2}, 4239 (1970).
\bibitem{cole_cohen} M.\,W.\,Cole and M.\,H.\,Cohen, Phys. Rev. Lett. {\bf 23}, 1238 (1969).
\bibitem{wigner} W.\,Wigner, Phys. Rev. {\bf 46}, 1002 (1934).
\bibitem{grimes_adams} C.\,C.\,Grimes, and G.\,Adams, Phys. Rev. Lett. {\bf 42}, 795 (1979).
\bibitem{platzman_dykman} P.\,M.\,Platzman and M.\,I.\,Dykman, Science {\bf 284}, 1967 (1999).
\bibitem{shikin} V.\,B.\,Shikin, Sov. Phys. JETP {\bf 31}, 936 (1970).
\bibitem{williams_crandall_willis} F.\,I.\,B.\,Williams, R.\,S.\,Crandall, and A.\,H.\,Willis, Phys. Rev. Lett., {\bf 26}, 7 (1971).
\bibitem{ando_fowler_stern} T.\,Ando, A.\,Fowler, and F.\,Stern, Rev. Mod. Phys. {\bf 54}, 437 (1982).
\bibitem{platzman_fukuyama} P.\,M.\,Platzman and H.\,Fukuyama, Phys. Rev. B {\bf 10}, 3150 (1974).
\bibitem{haque} M.\,Haque, I.\,Paul, and S.\,Pankov, Phys. Rev. B {\bf 68}, 045427 (2003).
\bibitem{leiderer1} P.\,Leiderer and M.\,Wanner, Phys. Lett. {\bf 73A}, 189 (1979).
\bibitem{ebner} W.\,Ebner and P.\,Leiderer, Phys. Lett. {\bf 80A}, 277 (1980)
\bibitem{zinn-justin} J.\,Zinn-Justin {\it Quantum field theory and critical phenomena} (Clarendon Press, 1996).
\bibitem{thouless} D. J.\,Thouless, Phys. Rep. {\bf 13}, 93 (1974).
\bibitem{datta} S.\,Datta, {\it Electronic transport in mesoscopic systems} (Cambridge University Press, 1995).
\bibitem{monarkha_kono} Yu.\,P.\,Monarkha and K.\,Kono, {\it Two-dimensional Coulomb liquids and solids} (Springer, 2004).
\bibitem{lev_zhugayevich} B.\,I.\,Lev and A.\,Y.\,Zhugaevych, Phys. Rev. E, {\bf 57}, 6460 (1998).
\bibitem{leiderer2} P.\,Leiderer {\it Two-Dimensional Electron Systems by Ed. E.Y. Andrei}, ( Springer Netherlands 1997).
\bibitem{shikin_leiderer} V.\,B.\,Shikin and P.\,Leiderer, Sov. Phys. JETP Lett., {\bf 54}, 92, (1981).
\bibitem{tsui_stormer_gossard} D.\,C.\,Tsui, H.\,L.\,Stormer, and A.\,C.\,Gossard, Phys. Rev. Lett. {\bf 48}, 1559 (1982).
\bibitem{laughlin} R.\,B.\,Laughlin, Phys. Rev. Lett. {\bf 50}, 1395 (1983).
\bibitem{GrainDynamics} B.\,I.\,Lev and A.\,G.\,Zagorodny, Phys. Lett. A, {\bf 373}, 1101 (2009).
\bibitem{StronglyCollisional} B.\,I.\,Lev, V.\,B.\,Tymchyshyn, and A.\,G.\,Zagorodny, Cond. Matter Phys. {\bf 12} (2009).
\bibitem{myPublication} B.\,I.\,Lev, V.\,B.\,Tymchyshyn, and A.\,G.\,Zagorodny, Phys. Lett. A, {\bf 375}, 593 (2011).
\bibitem{krasnoholovets_lev} V.\,Krasnoholovets and B.\,I.\,Lev, Cond. Matter Phys. {\bf 6}, 67 (2003).
\bibitem{gorkov_chernikova} L.\,P.\,Gor'kov and D.\,M.\,Chernikova, JETP Lett. {\bf 18}, 119 (1973).

\bibitem{archer} A.\,J. Archer, Phys. Rev. E {\bf 78}, 031402 (2008).
\bibitem{imperio_reatto} A.\,Imperio and L.\,Reatto, J. Chem. Phys. {\bf 124}, 164712 (2006). 
\bibitem{imperio_reatto2} A.\,Imperio and L.\,Reatto, Phys. Rev. E {\bf 76}, 040402 (2007).

\bibitem{edwards_lennard} S.\,F.\,Edwards and A.\,Lennard, J. Math. Phys. {\bf 3}, 778 (1962).
\bibitem{samuel} S.\,Samuel, Phys. Rev. D {\bf 18}, 1916 (1978).
\bibitem{hubbard} J.\,Hubbard, Phys. Rev. Lett. {\bf 3}, 77 (1959).
\bibitem{landau_lifshitz_8} L.\,D.\,Landau and E.\,M.\,Lifshitz, {\it Electrodynamics of continuous media} (Pergamon Press, 1984).
\bibitem{skachko} I.\,Skachko, {\it Phase diagram of a 2-dimensional electron system on the surface of liquid helium}, Ph.D. thesis, Rutgers, The State University of New Jersey (2006).
\bibitem{koutsoumpos} S.\,Koutsoumpos, {\it Surface State Electron Dynamics on Deformed Liquid Helium Films}, Ph.D. thesis, Konstanz (2010).
\bibitem{lambert} D.\,K.\,Lambert, {\it Electrons on the surface of liquid helium}, Ph.D. thesis, Lawrence Berkeley Laboratory (1979).
\bibitem{brown_grimes} T.\,R.\,Brown and C.\,C.\,Grimes, Phys. Rev. Lett. {\bf 29}, 1233 (1972).
\bibitem{kosterlitz_thouless} J.\,M.\,Kosterlitz and D.\,J.\,Thouless, J. Phys. C, 1181 (1973).
\bibitem{nesterov} Yu.\,Nesterov, {\it Introductory Lectures on Convex Optimization. A Basic Course.} (Kluwer Academic Publishers, 2004).



\end{thebibliography}
\end{document}